\documentclass[longauth]{aa}  % 2 columns format

\usepackage{graphicx}
\usepackage{txfonts}
\usepackage{hyperref}
\usepackage{natbib}
\usepackage{lscape}
\usepackage{pdflscape}
\usepackage{soul}

\hypersetup{
        colorlinks=true,    % false: boxed links; true: colored links
        linkcolor=blue,  % color of internal links
        citecolor=blue,    % color of links to bibliography
        filecolor=cyan,     % color of file links
        urlcolor=black,      % color of external links
} % backref must be defined when \usepackage

\begin{document} 

\title{Cosmic Vine: A z=3.44 large-scale structure hosting massive quiescent galaxies}

%\subtitle{XX}
\titlerunning{Massive Quiescent galaxies in the Cosmic Vine at redshift 3.44}
\authorrunning{Jin et al.}

\author{
Shuowen Jin\inst{1,2,\thanks{Marie Curie Fellow}},
Nikolaj B. Sillassen\inst{1,2},
Georgios E. Magdis\inst{1,2,3},
Malte Brinch\inst{1,2},
Marko Shuntov\inst{1,3},
Gabriel Brammer\inst{1,3},
Raphael Gobat\inst{4},
Francesco Valentino\inst{5,1},
Adam C. Carnall\inst{6},
Minju Lee \inst{1,2},
Aswin P. Vijayan\inst{1,2},
Steven Gillman\inst{1,2},
Vasily Kokorev\inst{7},
Aurélien Le Bail\inst{8},
Thomas R. Greve\inst{1,2},
Bitten Gullberg\inst{1,2},
Katriona M. L. Gould\inst{1,3},
and Sune Toft\inst{1,3}
  %\fnmsep\thanks{Just to show the usage of the elements in the author field}
          }

   \institute{Cosmic Dawn Center (DAWN), Denmark\\
      \email{shuji@dtu.dk, shuowen.jin@gmail.com}
    \and
            DTU Space, Technical University of Denmark, Elektrovej 327, DK-2800 Kgs. Lyngby, Denmark
    \and
            Niels Bohr Institute, University of Copenhagen, Jagtvej 128, DK-2200 Copenhagen, Denmark
    \and
            Instituto de Física, Pontificia Universidad Católica de Valparaíso, Casilla 4059, Valparaíso, Chile
    \and
            European Southern Observatory (ESO), Karl-Schwarzschild-Strasse 2, Garching 85748, Germany
    \and
            Institute for Astronomy, University of Edinburgh, Royal Observatory, Edinburgh, EH9 3HJ, UK
    \and
            Kapteyn Astronomical Institute, University of Groningen, PO Box 800, 9700 AV Groningen, The Netherlands
    \and    Department of Physics, University of California Merced, 5200 North Lake Road, Merced, CA 95343, USA}

   \date{Received XXX / Accepted XXX}

 \abstract{We report the discovery of a large-scale structure at $z=3.44$ revealed by JWST data in the Extended Groth Strip (EGS) field. This structure, called the  Cosmic Vine, consists of 20 galaxies with spectroscopic redshifts at $3.43<z<3.45$ and six galaxy overdensities ($4-7\sigma$) with consistent photometric redshifts, making up a vine-like structure extending over a $\sim4\times$0.2~pMpc$^2$ area. The two most massive galaxies ($M_*\approx10^{10.9}~M_\odot$) of the Cosmic Vine are found to be quiescent with bulge-dominated morphologies ($B/T>70\%$). Comparisons with simulations suggest that the Cosmic Vine would form a cluster with halo mass $M_{\rm halo}>10^{14}M_\odot$ at $z=0$, and the two massive galaxies are likely forming the brightest cluster galaxies (BCGs). The results unambiguously reveal that massive quiescent galaxies can form in growing large-scale structures at $z>3$, thus disfavoring the environmental quenching mechanisms that require a virialized cluster core. Instead, as suggested by the interacting and bulge-dominated morphologies, the two galaxies are likely quenched by merger-triggered starburst or active galactic nucleus (AGN) feedback before falling into a cluster core. Moreover, we found that the observed specific star formation rates of massive quiescent galaxies in $z>3$ dense environments are one to two orders of magnitude lower than that of the BCGs in the TNG300 simulation. This discrepancy potentially poses a challenge to the models of massive cluster galaxy formation. Future studies comparing a large sample with dedicated cluster simulations are required to solve the problem.}


\maketitle

%_______________________________________________________

\section{Introduction}

Galaxy clusters are the most massive gravitationally bound structures in the Universe. Brightest cluster galaxies (BCGs) are the most luminous and massive elliptical galaxies located at the centers of galaxy clusters.
Studying the progenitors of galaxy clusters and their BCGs in the early Universe is fundamental for our understanding of galaxy formation and evolution.
In the past decade, massive and dense structures of galaxies have been continuously discovered at high redshift from $z\sim2$ out to the epoch of reionization (e.g., \citealt{Capak2011Nature,Walter2012,Mei2015cluster,Wang_T2016cluster,Mantz2018XLSSC122,Oteo2018cluster,Miller2018cluster_z4,ZhouLu2023cluster,Brinch2023b,Morishita2023z8cluster}). These structures have large scales, some extend over tens to hundreds of comoving Mpcs (e.g., \citealt{Koyama2013cluster,Cucciati2018,Forrest2023LSS}), and most host a high abundance of  star-forming galaxies. In these structures the most massive members are usually rich in gas and dust, and show vigorous star formations and complex morphologies. Simulations suggest that some of them would collapse and form galaxy clusters at later cosmic time, and hence they are likely proto-clusters hosting proto-BCGs (e.g., \citealt{Chiang2013cluster,Chiang2017cluster,Rennehan2020simu,Ata2022Nature,Montenegro-Taborda2023BCG}).
However, when and how the proto-BCGs quenched their star formation and transformed their morphology remain open questions.

In the $z<1$ Universe it is well established that environmental quenching \citep{Peng_YJ2010} is the dominant channel ceasing star formation in cluster galaxies, where galaxies were quenched via gas stripping and strangulation after falling into a virialized cluster core \citep{Gunn1972,Larson1980,Moore1998,Laporte2013cluster,Peng2015Nature,Shimakawa2018SW,Boselli2022review}. 
This quenching process was often presumed for $z>2$ proto-clusters. For example, \cite{Shimakawa2018SW} proposed a scenario where the first generation of massive quiescent cluster galaxies is formed in an already collapsed cluster core where the environmental quenching is taking place.
Nonetheless, at $z>1$ this picture has been debated by multiple studies (e.g., \citealt{Gobat2013,van_der_Burg2013,van_der_Burg2020,Webb2020,Ahad2023quench}) that argue  that most massive cluster galaxies are quenched by self-driven processes (e.g., mass quenching, AGN feedback) before entering a cluster core.
Therefore, detailed study of high-redshift quiescent galaxies and their environments is crucial to disentangling the quenching mechanisms.

Recently, quiescent members have been spectroscopically identified in galaxy overdensities at $z\gtrsim3$ \citep{Kubo2021quiescent,Kubo2022quiescent,McConachie2022,Ito2023,Shi_DD2023,Sandles2023}, which are exquisite samples to test the environmental quenching models.
However, the shallow depth of photometric surveys and the high incompleteness of spectroscopy observations has hampered our efforts to reveal their large-scale structures and assess the dynamical status of their local environment;  it is unclear whether they are hosted by a virialized cluster core.
The situation is currently changing with the successful operation of the James Webb Space Telescope (JWST). Its unprecedented sensitivity and long wavelength coverage allow us to efficiently select distant quiescent galaxies and reveal their large-scale environments.

On the other hand, massive quiescent galaxies have been identified at $z>3$ (e.g., \citealt{Glazebrook2017Natur,Schreiber2018Jekyll,Forrest2020QG,Forrest2020b,D'Eugenio2021QG,Valentino2023quiescent,Carnall2023Nature}), but their large-scale environments are barely studied due to the lack of deep imaging and spectroscopy follow-ups on megaparsec scales.
Accordingly, a megaparsec-scale structure at $z>3$ hosting massive quiescent galaxies in a well-defined JWST survey field would be an ideal laboratory to study the quenching and formation of proto-BCGs.
In this paper we report a large-scale structure called the Cosmic Vine at $z=3.44$ in the Extended Groth Strip (EGS) field covered by JWST surveys, and investigate two massive galaxies in the structure.
We adopt flat $\Lambda$CDM cosmology with $H_0=70$~km~s$^{-1}$~Mpc$^{-1}$, $\Omega_M=0.3$, as well as a Chabrier initial mass function \citep{Chabrier2003}.

\begin{figure*}[ht]
\centering
\includegraphics[width=0.98\textwidth]{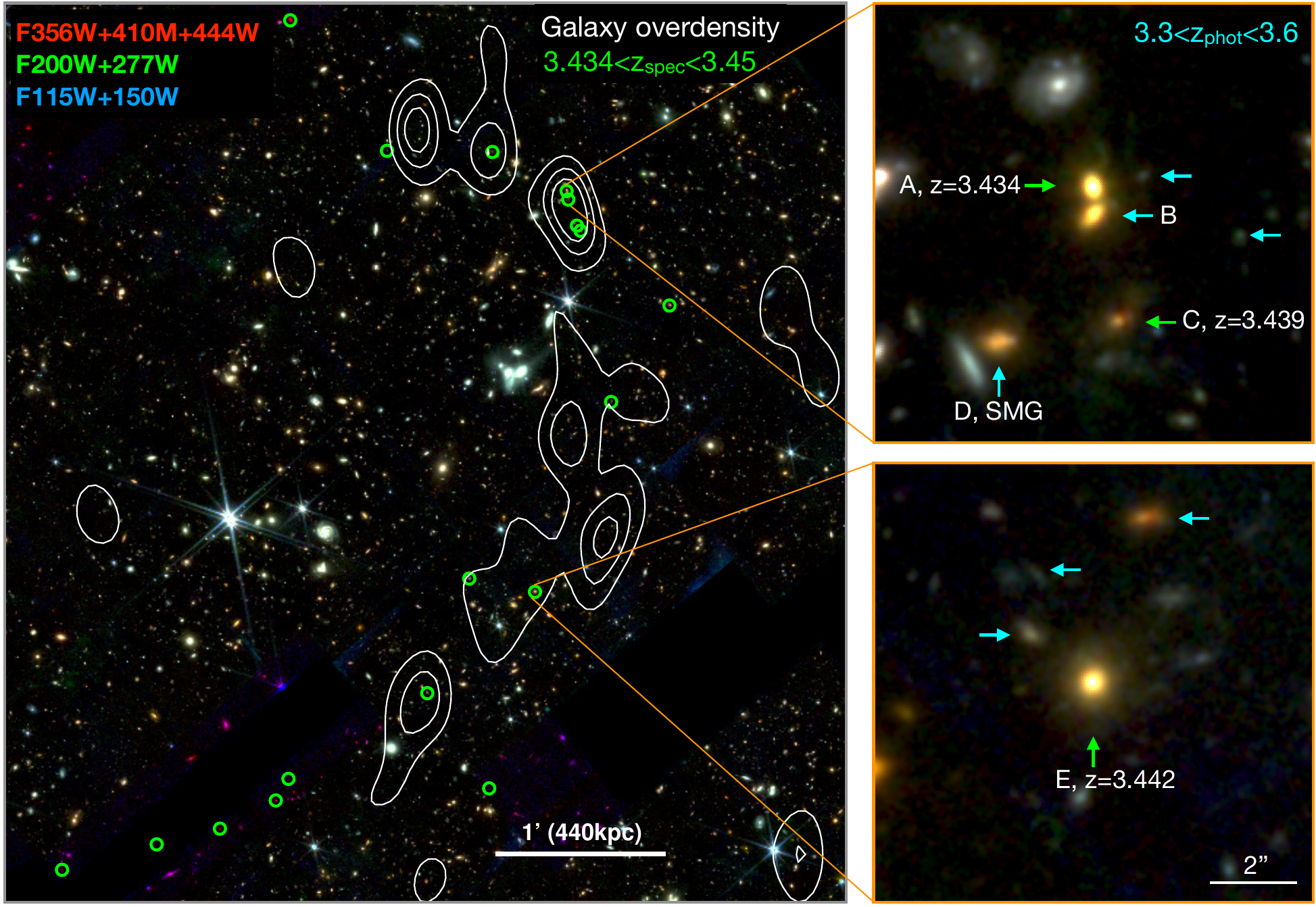}
\caption{JWST color-composed image of the Cosmic Vine (Red: F356W+F410W+F444W; Green: F200W+F277W; Blue: F115W+F150W). {\it Left:} The large scale structure. Sources with $3.435<z_{\rm spec}<3.455$ are marked with green circles. The white contours show the overdensity of $3.2<z_{\rm phot}<3.7$ sources in step levels of 2, 4, and 6$\sigma$. {\it Right:} $10''\times10''$ images centered on two massive galaxies in the Cosmic Vine. Galaxies with $z_{\rm spec}$ are highlighted with green arrows, and candidate members with $3.3<z_{\rm phot}<3.6$ are marked with cyan arrows.}
\label{fig:img}
\end{figure*}

\section{Data and methodology}

\subsection{Data processing and measurements}

This study used photometric data from JWST and the Hubble Space Telescope (HST), and spectroscopy data from JWST and the literature.
The JWST+HST photometric data and catalogs are publicly available in the Dawn JWST Archive (DJA),\footnote{https://dawn-cph.github.io/dja} and the reduced images and spectra have been visualized on the DJA Interactive Map Interface.\footnote{https://dawn-cph.github.io/dja/general/mapview}

The JWST imaging data are from the Cosmic Evolution Early Release Science survey (CEERS, \citealt{Finkelstein2023ceers}).
The data reduction, calibration, and source extraction follow the same pipeline applied in multiple studies (e.g., \citealt{Valentino2023quiescent,Jin2023cgg-z5,Gimenez-Arteaga2023,Kokorev2023dark,Gillman2023SMG}).
Briefly, we retrieved the pipeline-calibrated Stage 2 NIRCam products from the Mikulski Archive for Space Telescopes (MAST), then calibrated the data and processed them as mosaics using the \texttt{Grizli} package \citep{Brammer2021Grizli}.
The calibrated images are aligned to stars from the Gaia DR3 catalog \citep{Gaia2022}.
Sources were first extracted in the stacked map of long-wavelength (LW) images using source extraction and photometry (SEP, \citealt{Barbary2016SEP}), and photometry was measured within apertures of $0.3''$, $0.5''$, and $0.7''$ in diameter on the position from the extraction. 
The CEERS photometric catalog includes photometry of seven JWST bands (F115W, F150W, F200W, F277W, F356W, F410W, and F444W), and seven bands of HST (F105W, F125W, F140W, F160W, F435W, F606W, and F814W). 
We adopted $0.5''$ aperture photometry with aperture correction. 
The photometric redshifts were calculated using the \texttt{EAzY} code \citep{Brammer2008EAZY} that fit above photometry with a linear combination of 12 pre-selected flexible stellar population synthesis (FSPS) templates. 

The JWST spectroscopic observations used in this work are data from NIRSpec Prism grating (project ID: DD-2750, PI: P.
Arrabal Haro), which was taken using the Micro Shutter Assembly (MSA) multi-object spectroscopy (MOS) mode with ``clear'' filter. The data were reduced and calibrated using \texttt{MsaExp},\footnote{https://github.com/gbrammer/msaexp} following the reduction process described in \cite{Heintz2023spec}.
In short, we processed the spectroscopic data set using the custom-made pipeline \texttt{MsaExp} v. 0.6.7 \citep{Brammer2023MsaExp}. This code utilizes the Stage 2 products from the MAST JWST archive and performs standard calibrations for wavelength, flat-field, and photometry on the individual NIRSpec exposure files. 
MsaExp then corrects for the noise and the bias levels in individual exposures.
The 2D spectra are combined for individual exposures, and the 1D spectra are extracted using an inverse-weighted sum of the 2D spectra in the dispersion directions.
The NIRSpec Prism observations have wavelength coverage from 0.7$\mu$m to 5.3$\mu$m, with a varying spectral resolution from $R\sim50$ at the blue end
to $R\sim400$ at the red end.
The spectral redshifts are measured by fitting the 1D spectra with emission lines and continuum using \texttt{MsaExp}. We then visually inspected the fitted spectra and ranked the robustness of the redshift with grades from 0 to 3, which are 0$=$data quality problem; 1$=$no features; 2$=$with features but ambiguous redshift; 3$=$robust. We adopt the redshifts with robust features,  grade 3. Given the low resolution of NIRSpec Prism spectrum ($R\sim100$ at $3\mu$m), in Table~\ref{tab:1} we rounded the Prism redshifts to the precision of 0.001.

\subsection{Selection}

The structure, which we dub ``Cosmic Vine'', was initially selected by applying the overdensity mapping technique in \cite{Brinch2023cluster} with photometric redshifts $z_{\rm phot}$ from the CEERS catalog and seven spectroscopic redshifts $z_{\rm spec}$ from literature.
The overdensity mapping technique is based on a weighted adaptive kernel technique developed by \cite{Darvish2015} and \cite{Brinch2023cluster}.
In the overdensity mapping procedure, the photometric redshift uncertainties were accounted for in the weight of the chosen redshift bin, and spectroscopic redshifts $z_{\rm spec}$ are given with the highest weights. We performed the overdensity mapping in the  redshift range $2<z<5$ with redshift bin size of $5\%(1+z)$.
The Cosmic Vine was selected as the most significant overdensity in the redshift bin $3.29<z<3.77$ (Fig.~\ref{fig_selection}, top). As shown in Fig.~\ref{fig:img}, six peaks of galaxy overdensities are found with $>4\sigma$ significance over the field level, and three of them are found with $>6\sigma$. The primary overdensity peak (i.e., Peak A, upper right of Fig.~\ref{fig:img}) is centered on RA 214.86605, Dec 52.88426.
Subsequently, we also searched for extra spectroscopic redshifts in the latest DJA archive and the literature. As listed in Table~\ref{tab:1}, we found 20 spectroscopically confirmed members in total, whose   redshifts were collected from the DJA and multiple surveys \citep{Schreiber2018quiescent,Kriek2015MOSDEF,Urbano_Stawinski2023DEIMOS,Cooper2012DEEP3}. 
In Fig.~\ref{fig:img} we show 18 galaxies with $z_{\rm spec}$ in green circles; the other two sources (ID=42414, 46256 in Table~\ref{tab:1}) are located farther  north, and hence are not shown in the figure.
The two most massive members are Galaxy A and E (Fig.~\ref{fig:img}, right), and there are $\sim200$ candidate members with $3.3<z_{\rm phot}<3.6$, including a quiescent candidate (Galaxy B) selected by \cite{Valentino2023quiescent} and a submillimeter galaxy (Galaxy D) identified by \cite{Gillman2023SMG}.

\subsection{SED and spectral fitting}
\label{sed}

\begin{figure*}[ht]
\centering
\includegraphics[width=0.98\textwidth]{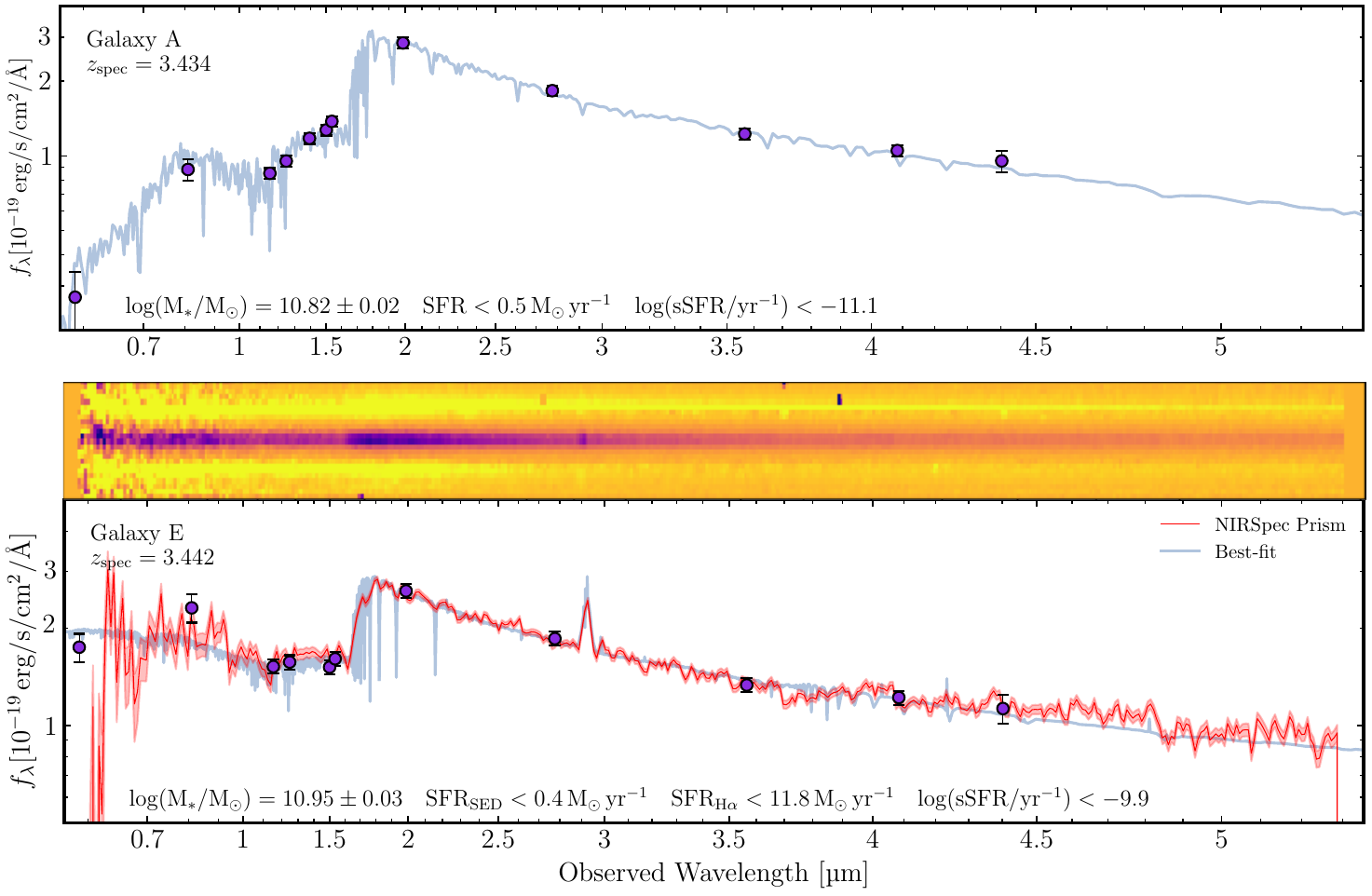}
\caption{SED and spectra of Galaxy A and E. The blue curves show the best fit of the Bagpipes fitting. For Galaxy E,  The NIRSpec Prism 2D spectrum is overplotted on the 1D spectrum. The 1D spectrum is shown in red with the uncertainty marked in shade. 
}
\label{fig:spec}
\end{figure*}

For the confirmed members, we fit the JWST+HST photometry and NIRCam spectra using the \texttt{Bagpipes} code \citep{Carnall2018Bagpipes} with fixed $z_{\rm spec}$. Following the recipes in \cite{Carnall2023Nature}, we assumed a double-power-law star formation history (SFH), the attenuation curve of \citet{Salim2018_dust_attenuation_curve}, and the radiation fields in the range of $-4<logU<-2$. We used a metallicity grid from log$(Z/Z_\odot)=-2.3$ to 0.70, $A_{V}$ grid from 0 to 4, and an age grid from 1~Myr to 2~Gyr.
For Galaxy A, we ran \texttt{Bagpipes} with its JWST+HST photometry at fixed $z_{\rm spec}=3.434$. As Galaxy E has NIRSpec Prism data and shows post-starburst features, we performed spectrophotometric fitting following a  method similar to that  in \cite{Strait2023PSB}. We first scaled the NIRSpec 1D spectra to the JWST photometry using a wavelength-dependent polynomial scaling curve. 
We note that Galaxy E shows a broad emission (${\rm FWHM}\sim3700$~km~s$^{-1}$) at the wavelength of H$\alpha$, which might be from active galactic nucleus (AGN) activity or a blending of H$\alpha$ and [NII], whereas it is not feasible to model broad H$\alpha$+[NII] with this low-resolution spectrum ($R\sim100$ at 3$\mu$m).
We thus fit a single Gaussian to the broad H$\alpha$ line and subtracted the best fit (FWHM$=3696$~km~s$^{-1}$) from the spectrum. 
%As narrow lines could be contaminated by AGN, 
Following \citep{Carnall2023Nature}, we masked out narrow emission lines and fit the broadline-subtracted and masked spectrum together with the JWST+HST photometry. As the continuum can be boosted by AGN and nebular, we included an AGN component and a nebular model to account for continuum emission from AGN and star-forming regions.
The best-fit results are presented in Fig.~\ref{fig:spec} and Table~\ref{tab:1}.

For the other confirmed members, we ran SED fitting with JWST+HST photometry using the same \texttt{Bagpipes} setups. As flagged by asterisks in Table~\ref{tab:1}, four sources are on the edge of or are out of the CEERS NIRCam mosaics. Two of them have NIRSpec spectroscopy, and we thus fit their spectra to derive stellar masses and star formation rates (SFRs). The other one has no JWST data, and we thus adopted the measurements from the EGS-CANDELS catalog \citep{Stefanon2017EGS}. The last source (ID=49474) is a Ly$\alpha$ emitter only found in the catalog of \cite{Urbano_Stawinski2023DEIMOS}, and thus no photometry is available for SED fitting.

\subsection{Morphology analysis}
\label{morphology}

In order to quantify the morphology, we used \texttt{SourceXtractor++} \citep{bertin20, kummel20} to fit the light profile of the JWST images over the whole CEERS survey field. 
To have meaningful results, the morphological fitting was only done for sources detected in F444W with ${\rm S/N}>20$.
For each source  we applied two models: a single-S\'ersic model with index varying from $n=1$ to 8 and a  Bulge+Disk decomposition with fixed index $n=4$ for bulge and $n=1$ for disk.
The single-S\'ersic fitting was performed by simultaneously fitting all available JWST images, and the Bulge+Disk decomposition was done for image of each JWST band.
In Fig.~\ref{fig_morph} we show an example of Bulge+Disk decomposition in F200W. The results of the single-S\'ersic index, effective radius, and bulge-to-total ratio ($B/T$) are listed in Table~\ref{tab:1} for the confirmed members of the Cosmic Vine. The error bars of the morphological parameters were obtained from the covariance matrix of the model fit, which was computed by inverting the approximate Hessian matrix of the loss function at the best-fit values. These error bars are found to be considerably underestimated by a factor of $2-3$ \citep[Shuntov et al. in prep.]{HubertEMC2022, MerlinEMC2023}, and should be considered only as a lower limit.

%\begin{landscape}
\begin{table*}
{
\caption{Confirmed members of the Cosmic Vine.}
\label{tab:1}
\renewcommand\arraystretch{1.2}
\centering
\begin{tabular}{ccccccccc}
\hline\hline
  ID & RA, Dec & $z_{\rm spec}$ & log($M_*/M_\odot$) & SFR & $r_{\rm eff}$ & $n$ & $B/T$ & Type\\
   &   (J2000) & & & [$M_\odot$~yr$^{-1}$] & [$''$] & & &\\
\hline
  56033 (A) & 214.86605,52.88426 & 3.434$^a$ & $10.82^{+0.02}_{-0.02}$ & $<0.5$ & $0.078\pm0.001$ & 2.66$\pm$0.01 & 0.73$\pm$0.01 & QG\\
  39138 (E) & 214.87123,52.84507 & 3.442$^b$ & $10.95^{+0.03}_{-0.03}$ & $<0.4$ & $0.215\pm0.001$ & 3.81$\pm$0.02 & 0.76$\pm$0.01 & QG \\
\hline
  2342 & 214.94776,52.81789 & 3.4360$^c$ & $10.55^{+0.16}_{-0.15}$ & $131^{+84}_{-40}$ &  -- & -- & -- &  SF\\
  12903$^*$ & 214.92221,52.82193 & 3.445$^b$  & $10.36^{+0.06}_{-0.05}$ & $293^{+38}_{-38}$ &  $0.599\pm0.004$ & 0.59$\pm$0.01 & 0.11$\pm$0.18 &  SF \\
  17600$^*$ & 214.91318,52.82468 & 3.4379$^c$ & $10.9^{+0.09}_{-0.10}$ & $15^{+4}_{-1}$ & $0.203\pm0.003$ & 3.97$\pm$0.06 & 0.11$\pm$0.17 &  SF\\
  19339 & 214.91113,52.82679 & 3.440$^b$ & $10.19^{+0.11}_{-0.17}$ & $15^{+3}_{-2}$ & -- &  -- & -- &  SF\\
  29557 & 214.87865,52.82586 & 3.4406$^d$ & $9.59^{+0.17}_{-0.21}$ & $6^{+1}_{-1}$ & $0.122\pm0.001$ & 1.87$\pm$0.10 & 0.11$\pm$0.21 &  SF\\
  30531 & 214.88879,52.83527 & 3.4375$^c$ & $8.66^{+0.08}_{-0.07}$ & $5^{+1}_{-1}$ & $0.242\pm0.006$ &  2.83$\pm$0.09 & 0.09$\pm$0.01 &  SF\\
  36814 & 214.88185,52.84637 & 3.435$^b$ & $9.66^{+0.10}_{-0.10}$ & $28^{+5}_{-5}$ & $0.454\pm0.002$ &  0.93$\pm$0.01 & 0.03$\pm$0.01 &  SF\\
  42414 & 214.96231,52.92031 & 3.4387$^d$  & $8.37^{+0.13}_{-0.36}$ & $2^{+1}_{-1}$ & $0.142\pm0.002$ &  0.30$\pm$.01 & 0.00$\pm$0.01 & SF \\
  46256 & 214.93161,52.90870 & 3.436$^b$ & $10.41^{+0.05}_{-0.06}$ & $213^{+26}_{-15}$ &  $0.016\pm0.001$ & 8.00$\pm$0.01 & 0.85$\pm$0.01 & AGN\\
  48525 & 214.89515,52.88820 & 3.450$^b$ & $8.57^{+0.08}_{-0.11}$ & $2^{+1}_{-1}$ & $0.047\pm0.001$ & 1.53$\pm$0.06 & 0.31$\pm$0.06 & SF \\
  49364 & 214.85886,52.86363 & 3.4365$^e$ & $9.57^{+0.07}_{-0.13}$ & $25^{+3}_{-2}$ &  $0.211\pm0.001$ & 0.43$\pm$0.01 & $<$0.01 & SF \\
  49474$^*$ & 214.91082,52.90096 & 3.4405$^{d,\dag}$ & -- & -- & -- & -- & -- & SF \\
  54034 & 214.87806,52.88808 & 3.449$^b$ & $9.18^{+0.07}_{-0.07}$ & $11^{+1}_{-1}$ &  $0.128\pm0.001$ & 1.83$\pm$0.03 & 0.61$\pm$0.01 & SF \\
  55035 & 214.86438,52.88086 & 3.4431$^c$ & $8.62^{+0.09}_{-0.08}$ & $4^{+1}_{-1}$ & $0.091\pm0.001$ &  1.22$\pm$0.03 & 0.04$\pm$0.02 & SF \\
  55136 & 214.86379,52.88037 & 3.4406$^f$ & $8.97^{+0.07}_{-0.06}$ & $10^{+1}_{-1}$ & $0.084\pm0.001$ & 3.77$\pm$0.05 & 0.70$\pm$0.01 & SF \\
  55785 (C) & 214.86578,52.88342 & 3.439$^b$ & $10.07^{+0.09}_{-0.13}$ & $19^{+16}_{-5}$ & $0.346\pm0.006$ & 3.39$\pm$0.06 & 0.57$\pm$0.01 & SF \\
  56241 & 214.84940,52.87305 & 3.4526$^e$ & $9.80^{+0.13}_{-0.09}$ & $72^{+25}_{-14}$ & $0.221\pm0.001$ & 1.75$\pm$0.01 & 0.26$\pm$0.01 & SF \\
  1345-11017$^*$ & 214.93245,52.82039 & 3.445$^b$ & $8.95^{+0.04}_{-0.03}$ & $10^{+1}_{-1}$ & -- & -- & -- & SF \\
\hline
\end{tabular}\\
}
{Notes: $^*$Sources on the edge of the  NIRCam LW mosaics; $^\dag$Uncertain redshift due to inconsistent $z_{\rm phot}=2.43^{+0.20}_{-0.14}$; $^a$Keck/MOSFIRE \citep{Schreiber2018quiescent}, $^b$JWST/NIRSpec (this work), $^c$MOSDEF \citep{Kriek2015MOSDEF}, $^d$Keck/DEIMOS \citep{Urbano_Stawinski2023DEIMOS}, $^e$Keck/MOSFIRE (this work), $^f$DEEP3 \citep{Cooper2012DEEP3}; $n$: S\'ersic index; $B/T$: Bulge-to-total ratio in F277W; Type: QG (quiescent galaxy), SF (star-forming). 
} 
\end{table*}
%\end{landscape}

\section{Results}
\subsection{Cosmic Vine: A large-scale structure at $z=3.44$}

In addition to the overdensity of photometric redshifts, 20 galaxies have been found with $3.434<z_{\rm spec}<3.45$ in the Cosmic Vine area. As listed in Table~\ref{tab:1}, the redshifts are confirmed by the JWST/NIRSpec spectroscopy, the Keck/MOSFIRE observations \citep{Schreiber2018quiescent,Kriek2015MOSDEF}, the Keck/DEIMOS survey of Ly$\alpha$ emitters \citep{Urbano_Stawinski2023DEIMOS}, and the DEEP3 survey \citep{Cooper2012DEEP3}. 
The galaxies with $z_{\rm spec}$ are shown as green circles in Fig.~\ref{fig:img}, which overlap well on the galaxy overdensities of photomeric redshifts.
The source with $z_{\rm spec}$ in the Cosmic Vine also dominates the available $z_{\rm spec}$ at $z\sim3.4$ in the $\sim100$~arcmin$^2$ CEERS field (Fig.~\ref{fig_selection}, bottom);  the overdensity of the  $z_{\rm spec}$ sources in the Cosmic Vine is $8.8\sigma$ above the field level.
This thus solidly confirms that the Cosmic Vine is a real structure at $z\sim3.44$.
Remarkably, the shape of the Cosmic Vine is significantly elongated, extends over a length of $\sim4$~Mpc, and has a narrow width of $\sim0.2~$pMpc on the sky, which is significantly larger than compact galaxy groups and proto-clusters at $z>3$ (e.g., \citealt{Oteo2018cluster,Miller2018cluster_z4,Daddi2022Lya,Sillassen2022HPC1001,ZhouLu2023cluster}). 
In the literature there are two structures that are very similar to the Cosmic Vine. The first is the $z\sim3.35$ large-scale structure PCl J0959+0235 reported by \cite{Forrest2023LSS}, which is at a similar redshift, and hosts multiple overdensity peaks on a similar scale and massive quiescent members \citep{McConachie2022}. The second    is the $z=2.2$ large-scale structure found by \cite{Spitler2012LSS}, which has a comparably long and vine-like shape.

We note that a ``tail'' made of five sources with $z_{\rm spec}$ is present on the bottom left of Fig.~\ref{fig:img}, but no galaxy overdensity has been found on it  because the five galaxies are on the edge of or are out of the CEERS NIRCam mosaics, and the photometric information is incomplete. 
Hence, the membership identification is limited by the area of the CEERS survey; the actual size of the Cosmic Vine would be larger if there were members that existed outside of the JWST mosaics.

\subsection{Massive quiescent galaxies}

Remarkably, the two most massive galaxies in the Cosmic Vine, Galaxy A and Galaxy E (Fig.~\ref{fig:img}), are found to be quiescent.
Galaxy A is located in the densest region of the Cosmic Vine, which is known as  Peak A. Galaxy A has been  classified as a quiescent galaxy in multiple studies \citep{Schreiber2018quiescent,Valentino2023quiescent,Carnall2023sample}, and was first  reported at $z_{\rm spec}=3.434$ by \cite{Schreiber2018quiescent} using Keck/MOSFIRE spectroscopy. Notably, the redshift $z=3.434$ has a confidence probability of 84\% and was flagged as an uncertain redshift in \cite{Schreiber2018quiescent}. However, using the latest JWST and HST photometry, the photometric redshift of Galaxy A has been constrained to be $z_{\rm phot}=3.53^{+0.08}_{-0.10}$ (16th, 84th quartiles) by EAzY SED fitting in the DJA catalog. As an independent measure, \cite{Carnall2023sample} estimated a $z_{\rm phot}=3.44^{+0.14}_{-0.08}$ using \texttt{Bagpipes} with a different version of JWST+HST photometry. The two $z_{\rm phot}$ results agree well with the $z_{\rm spec}=3.434$, and are consistent with the median redshift of Cosmic Vine within the $z_{\rm phot}$ uncertainty. Furthermore, the $z_{\rm phot}$ uncertainty of Galaxy A ($\Delta z\sim0.1$) is two times smaller than the median $z_{\rm phot}$ error of the other confirmed members. All these pieces of evidence support that Galaxy A is a member of the  Cosmic Vine.

With state-of-the-art JWST and HST photometry, as shown in Fig.~\ref{fig:spec}, the \texttt{Bagpipes} SED fitting yields a stellar mass of log$(M_*/M_{\odot})=10.82\pm0.02$ and an upper limit of SFR$<0.5~M_{\odot}$~yr$^{-1}$ (95th quantile), confirming its massive and quiescent nature. The inferred SFH suggests a post-starburst picture with a peak of star formation 350~$M_\odot$~yr$^{-1}$ occurring at $z\sim4.5$ and being quiescent by $z=4$ (Fig.~\ref{fig_SFH}).
The peak SFR is comparable with that of submillimeter galaxies (SMGs) at $z\sim4$ (e.g., \citealt{Jin2022ColdDustyGuys}).
Coincidentally, a tidal tail associated with Galaxy A is robustly detected in NIRCam F200W and LW images (Fig.~\ref{fig_morph}), indicating a merger morphology. Our morphology analysis gives a S\'ersic index of $n\sim2.7$ and a  bulge-to-total ratio of $B/T>0.7$, revealing a bulge-dominated morphology. Moreover, the size of Galaxy A is extremely compact with an effective radius of $r_{\rm eff}=622\pm3$~pc. The size and the stellar mass surface density within the $r_{\rm eff}$ (log$(\Sigma_{\rm eff})=10.43\pm0.03~M_\odot$~kpc$^{-2}$) are comparable with that of compact starburst galaxies \citep{Puglisi2019size,Gullberg2019SMG,Diamond-Stanic2021}, which again supports the major merger and post-starburst nature.

Galaxy E was selected as a quiescent candidate by \cite{Merlin2019passive} in the \cite{Stefanon2017EGS} catalog. Recently, it was re-selected by \cite{Carnall2023sample} using its specific star formation rate (sSFR) derived from SED fitting with JWST photometry ($z_{\rm phot}=3.53\pm0.12$), and also selected by \cite{Valentino2023quiescent} using the $NUVUVJ$ diagram in \cite{Gould2023quiescent}. It is the most massive galaxy in the Cosmic Vine with a log$(M_*/M_\odot)=10.95\pm0.03$, which is confirmed at $z=3.442$ with JWST/NIRSpec Prism spectroscopy (Fig.~\ref{fig:spec}). 
Galaxy E is well detected with a H$\alpha$ emission and a strong Balmer break. As no other lines are present in the spectrum, it appears to be a post-starburst galaxy (e.g., \citealt{Chen_YM2019PSB,French2021PSB}).
The \texttt{Bagpipes} fitting of the NIRSpec spectrum shows negligible star formation with an upper limit of SFR$_{\rm SED}<0.4~M_\odot$~yr$^{-1}$ (95th quantile) and a moderate attenuation ${A_V=0.45\pm0.06}$. The inferred SFH is relatively uncertain, but suggests a quenching time at $z\sim4$.

The H$\alpha$ emission of Galaxy E appears dominated by a broad component (FWHM$=3696\pm324$~km~s$^{-1}$), which suggests AGN activity or blending of [NII]+H$\alpha$. Given the low resolution of the prism spectrum, the two cases cannot be distinguished with current data, and high-resolution spectroscopy is required to identify the potential AGN activity. However, here we derived a SFR$_{\rm H\alpha}$ upper limit for the two cases. For the first, assuming the broad component is from an AGN, the residual is minimal after subtracting the best-fit broad Gaussian (i.e.,  $1.39\times10^{-18}$~erg/s/cm$^{2}$). 
By integrating the H$\alpha$ absorption of the best-fit model, we obtained an upper limit for narrow H$\alpha$ flux of $3.45\times10^{-18}$~erg/s/cm$^{2}$. Accounting for the attenuation, it gives a constraint of ${\rm SFR_{H\alpha}< 1.8~M_\odot}$~yr$^{-1}$ according to the H$\alpha$-SFR correlation in \citet{Pflamm-Altenburg2007_Ha_SFR}.
This might suggest that Galaxy E is a quiescent galaxy hosting an active black hole, similar to the $z=4.7$ GS-9209 \citep{Carnall2023Nature}.
For the second case, assuming there is no any AGN contribution to the H$\alpha$ emission, the integrated H$\alpha$ flux would be $2.59\times10^{-17}$~erg/s/cm$^{2}$. Adopting a ratio of [NII]/H$\alpha=0.3$, which is  typical for star-forming galaxies, we obtained an upper limit of ${\rm SFR_{H\alpha}<11.8}~M_\odot$~yr$^{-1}$. We note that this is a conservative limit because the [NII]/H$\alpha$ ratio can be high in high-z quiescent galaxies (e.g., [NII]/H$\alpha=0.97$ in \citealt{Carnall2023Nature}), and the H$\alpha$ from star formation could be even fainter if there is any AGN activity. 
The two SFR limits give sSFR upper limits of log(sSFR/yr$^{-1}$)$<-10.7$ and log(sSFR/yr$^{-1}$)$<-9.9$, respectively. Both results are compatible with the sSFR from the SED fitting (Fig.~\ref{fig_SFH}), and support the quiescent nature of Galaxy E. Here we adopt the more conservative limit of ${\rm log(sSFR/yr^{-1})<-9.9}$. 

Our morphology analysis shows that Galaxy E has a S\'ersic index of 3.81 that is close to local elliptical galaxies, and the bulge--disk decomposition gives a $B/T=0.76$, revealing a bulge-dominated morphology. 
In contrast to Galaxy A, Galaxy E is located in a relatively isolated environment, where the local overdensity is just above the field level with a 2$\sigma$ significance. No robust interacting features are found on Galaxy E, and its effective radius is about three times larger than that of Galaxy A.
In comparison with the $z\sim0.1$ post-starburst galaxies that have an average $n=1.7$ \citep{Sazonova2021PSB}, the S\'ersic indices of the two galaxies are larger by a factor of 1.6 and 2.2, respectively.

We note that the SED of Galaxy E is bluer than Galaxy A and other typical quiescent galaxies, which occurs  because   the blue part of our best-fit model ($\lambda_{\rm obs}<1.5~\mu$m) is dominated by AGN. SED fitting without an AGN would yield a high SFR$=495\pm102~M_\odot$~yr$^{-1}$ with high attenuation $A_V=1.00
\pm0.04$. With such a  high SFR and attenuation Galaxy E would be detected in the far-infrared (FIR) and (sub)millimeter. 
We checked ancillary FIR and millimeter data sets (MIPS, Herschel, and SCUBA2), and found that Galaxy E is not detected in any images.
Furthermore, we made use of the Super-deblended FIR+submm+radio catalog in the EGS field from Le Bail et al. (in prep.), in which they deblended the MIPS, Herschel, SCUBA2, and AzTEC images using the Super-deblending technique \citep{Jin2018cosmos,Liu_DZ2017}. As in Fig.~\ref{fig_FIR}, Galaxy E is not detected in any FIR or (sub)millimeter bands, and is only tentatively detected at MIPS 24$\mu m$ and VLA 3GHz with ${\rm S/N\sim3}$. We performed a panchromatic NIR-to-radio SED fitting and obtained an upper limit of ${\rm SFR_{IR}<210~M_\odot~yr^{-1}}$. This FIR SFR is compatible with the ${\rm SFR_{H\alpha}}$ limit, but disagrees with the dusty ${\rm SFR\approx500~M_\odot~yr^{-1}}$ solution, and hence the dusty star-forming scenario is disfavored for Galaxy E.

\begin{figure*}[h]%
\centering
\includegraphics[width=0.48\textwidth]{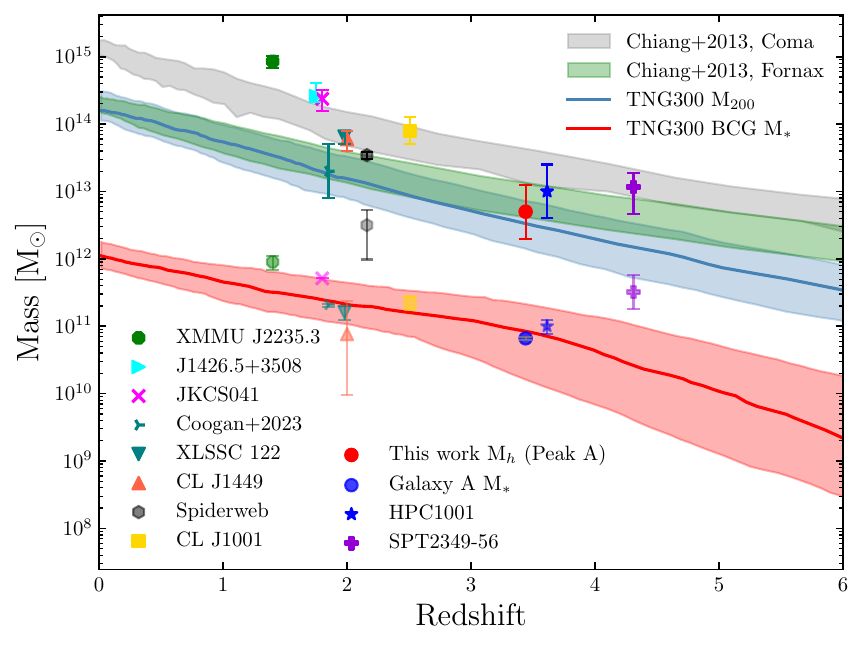}
\includegraphics[width=0.48\textwidth]{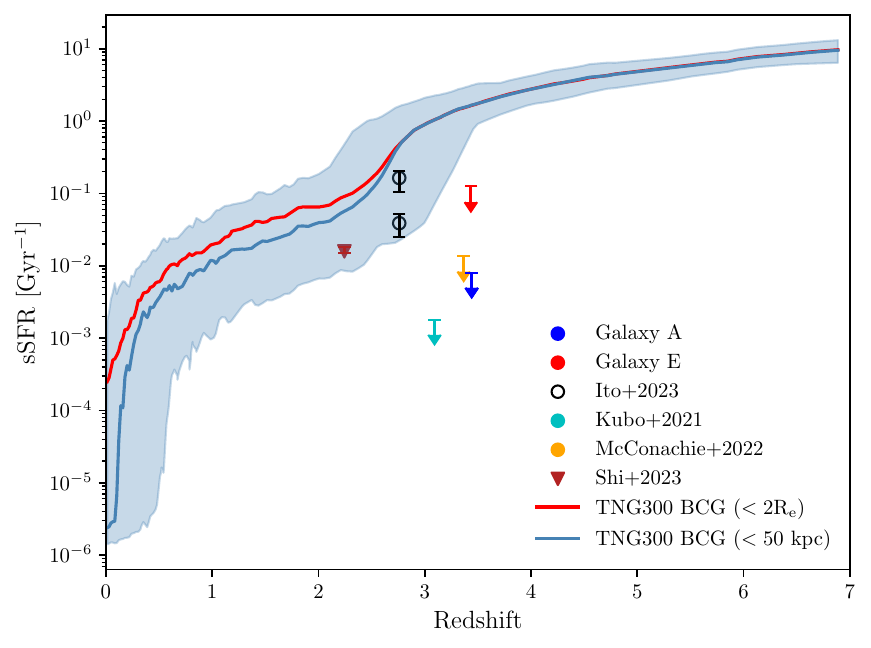}
\caption{Comparison with simulations.
 {\it Left:} Mass vs. redshift for literature proto-clusters and TNG300 simulations.
The gray and blue shaded areas show the simulated halo mass evolution of proto-clusters in the  \cite{Chiang2013cluster} and TNG300 simulations \citep{Montenegro-Taborda2023BCG}, repectively. The red shaded area marks the stellar mass of BCGs in TNG300 \citep{Montenegro-Taborda2023BCG}. The halo mass of peak A and the stellar mass of Galaxy A are consistent with the predictions from the models, suggesting a massive descendant with a halo mass of more than $10^{14}~M_\odot$ at $z=0$.
{\it Right:} sSFR vs. redshift for BCGs. The sSFRs of BCGs in TNG300 simulations \citep{Montenegro-Taborda2023BCG} are shown as blue and red curves, overlaying with the observed sSFRs of massive quiescent members in $z>2$ proto-clusters. The blue shaded regions indicate the 16th to 84th percentile range of the $r<50$~kpc sSFR measurements in TNG300.
}
\label{fig:simu}
\end{figure*}

\subsection{Halo mass}

As the Cosmic Vine is an extremely long and large structure ($\sim4$~pMpc), it is unlikely to be hosted by a single dark matter halo. Although, the densest region Peak A might be already collapsed. 
We thus estimated the dark matter halo mass of the Peak A following the methods in \cite{Sillassen2022HPC1001}: (1) Using the $M_{\rm halo}$-$M_{\ast}$ scaling relation from \cite{Behroozi2013Mhalo} and the stellar mass of Galaxy A, it yields a halo mass of log$(M_{\rm halo}/M_\odot)=12.5\pm0.4$; (2) We obtained a total stellar mass of $M_{\rm *,total}=(2.6\pm0.4)\times10^{11}~M_\odot$ by summing the stellar masses down to $10^7M_{\odot}$ of all confirmed and candidate members in the Peak A within a radius of 15" (111~pkpc). Adopting the dynamical mass-constrained $M_{\rm halo} - M_{\ast}$ scaling relation for $z\sim1$ clusters with $0.6 \times 10^{14} <M/M_{\odot}<16\times 10^{14}$ \citep{van_der_Burg2014} yields
a halo mass of $\log(M_{200}/M_{\odot})=12.8$; 
(3) Adopting the stellar-to-halo mass relation of \cite{Shuntov2022Mhalo} and $M_{\rm *,total}=(2.6\pm0.4)\times10^{11}~M_\odot$, we obtained a halo mass of log$(M_{\rm halo}/M_\odot)=12.7$; 
(4) Assuming a group velocity dispersion $\sigma_V=400$\,km~s$^{-1}$, we found that the galaxy number of Peak A (in a putative $R_{\rm vir}<15''$) is more overdense than the average field density by a factor of 97 at $z\sim3.4$ in the CEERS catalog. Applying a mean baryon and dark matter density of $7.41\times10^{-26}~$kg~m$^{-3}$ in comoving volume and a galaxy bias factor of 10--20 at $z=3.4$ \citep{Tinker2010haloBias}, we obtained a halo mass of log$(M_{\rm halo}/M_\odot)=12.4-12.7$.
The four methods agree on an average log$(M_{\rm halo}/M_\odot)=12.66$ with a scatter of 0.26 dex. We adopted a halo mass of log$(M_{\rm halo}/M_\odot)=12.7$ with a conservative uncertainty of 0.4 dex that is representative at these faint levels (e.g., \citealt{Daddi2021Lya,Sillassen2022HPC1001}).

\section{Discussion}

\subsection{Quenching mechanisms}

The elongated shape, the large size ($\sim4$~pMpc), and wide velocity range ($\sim1100$~km~s$^{-1}$) suggest that the Cosmic Vine is not a virialized system.
The abundance of star-forming galaxies (Table~\ref{tab:1}), a confirmed member of Type 1 AGN (Table 1, ID=46256), and a potential SMG cluster member  (Fig.~\ref{fig:img}, source D; see also \citealt{Gillman2023SMG}) indicate that the cluster is in its growing phase \citep{Shimakawa2018SW}.
In comparison with the $z=2.16$ Spiderweb proto-cluster \citep{Koyama2013cluster,Shimakawa2018SW,Jin2021cluster}, the Cosmic Vine has a comparable co-moving size and velocity range. 
However, the Spiderweb is at least partially virialized, as is evident from the extended X-ray emission and the detection of the  Sunyaev–Zeldovich effect \citep{Tozzi2022Spiderweb,Tozzi2022diffuseX,Di_Mascolo2023Natur}. 
In contrast,  Peak A of the Cosmic Vine is approximately eight times  less massive than the core of Spiderweb, and the projected shape appears elongated, which means that it is unlikely a virialized structure.
Surprisingly, two massive quiescent galaxies   formed in this large structure, in which Galaxy E is explicitly not in the core region but already quenched. 
This indicates that a cluster core is not essential for quenching massive cluster galaxies, and quenching mechanisms that require a virialized cluster core are thus disfavored. 
We realized that recent studies suggest that ram-pressure stripping (RPS) can occur in local clusters that are not fully virialized (e.g., \citealt{Lourenco2023RPS}), where hot intercluster medium (ICM) has formed in clusters with log$(M_{\rm halo}/M_\odot)=14-15$ that are undergoing merging.
However, the core of the Cosmic Vine is less massive in $M_{\rm halo}$ by more than one order of magnitude, and a  hot ICM is unlikely to form.
On the other hand, RPS is expected to suppress star formation in low-mass galaxies more efficiently than in massive galaxies. On the contrary, in the Cosmic Vine only the most massive members are quenched.
For example, Galaxy C is spectroscopically confirmed at $z=3.439$ and located well in the core region. It is less massive than Galaxy A by a factor of six, but it is  fairly star-forming, which is inconsistent with the picture of gas stripping.

The challenge is to determine what culprit is quenching their star formations at such an early cosmic time.
Thanks to the high sensitivity and the long-wavelength coverage of JWST, the two quiescent galaxies are revealed with interesting features that allow us to assess their quenching mechanisms. 
As shown in Table 1, the two galaxies show bulge-dominated morphologies ($B/T>0.7$). Galaxy A has an extremely compact bulge and a tidal tail, both of which  point to a merger. Galaxy E shows potential AGN acitivity. 
On the other hand, SFHs from SED fitting suggest they were quenched at $4<z<6$. 
Given that the post-merger timescale is $\sim1$~Gyr \citep{Lotz2008merger}, this allows the merger event to happen before the starburst and quenching phases, as suggested by the SFHs.
Therefore, it is likely that the two galaxies were quenched by merger-triggered starbursts in the past 500~Myr. 
Strong AGN feedback is also a possible  quench to star formation; however, this is  difficult to verify since the AGN activity could  have taken place after the quenching of the galaxy ($z<4$).

\subsection{Comparison with simulations}

We compare the halo mass estimate with the masses of proto-clusters in \cite{Chiang2013cluster} and TNG300 simulations \citep{Montenegro-Taborda2023BCG}.
\cite{Chiang2013cluster} used a semi-analytical galaxy formation model \citep{Guo2011simu} run on the dark matter-only N-body simulation Millennium \citep{Springel2005Natur}, with which they tracked the evolution of dark matter and galaxies in about 3000 clusters from $z=7$ to $z=0$.
\cite{Montenegro-Taborda2023BCG} selected 280 systems with $M_{200}\geq10^{14}~M_\odot$ at $z=0$ in the TNG300 simulation \citep{Pillepich2018TNG} and traced their progenitors and proto-BCGs at high redshift.
In Fig.~\ref{fig:simu} we compare  the halo mass of Peak A with the results from the  \cite{Chiang2013cluster} model and TNG300. As Galaxy A is likely a proto-BCG, we also compare  its stellar mass with that of BCGs in TNG300 \citep{Montenegro-Taborda2023BCG}.
We found that the halo mass of Peak A is consistent with the  progenitor of a Fournax-class cluster in the TNG300 simulations, even accounting for the halo mass uncertainty, and it also partially agrees with the theoretical prediction in \cite{Chiang2013cluster}. This suggests that   Peak A would evolve to a cluster with $M_{\rm halo}>10^{14}$ at $z=0$. 
Although the halo mass of Peak A might be lower if it is not virialized, merging with nearby overdensities at a later time would significantly increase the mass to above that of  the cluster progenitors, and the final mass can be even more massive if galaxies in the large scale fall into  Peak A (e.g., \citealt{Ata2022Nature}).
Meanwhile, the stellar mass of  Galaxy A is also consistent with the BCG progenitors at $z=3.44$ (Fig.~\ref{fig:simu}, left), supporting the idea that Galaxy A is a proto-BCG. The consistency with simulations supports the idea that the Cosmic Vine is on the way to forming a cluster.
Furthermore, the halo and stellar masses of other massive proto-clusters at $1.3<z<4.4$ are also found to be  consistent with the simulations (e.g., \citealt{Rosati2009cluster,Stanford2012cluster,Gobat2013,Andreon2014cluster,Mantz2018XLSSC122,Wang_T2016cluster,Miller2018cluster_z4,Sillassen2022HPC1001,Coogan2023ICL,Shimakawa2023Spiderweb,Perez-Martinez2023_Spiderweb,Tozzi2022diffuseX,Di_Mascolo2023Natur}), which again supports the picture of a forming cluster. 
We note that Galaxy E has a slightly higher stellar mass than Galaxy A, but is relatively isolated;   it is possible that Galaxy E will become a BCG if it falls into the cluster core at a later cosmic time.

Moreover, we compared the sSFRs of $z>2.5$ quiescent cluster galaxies in the literature \citep{Kubo2021quiescent,Kubo2022quiescent,McConachie2022,Ito2023,Shi_DD2023} to the BCGs in TNG300 \citep{Montenegro-Taborda2023BCG}. 
The sSFRs of BCGs in TNG300 were measured within the radius of $2R_e$ and $50$~kpc, respectively.
We found that the observed sSFRs at $z>3$ are all lower than the predictions from TNG300 by one to two orders of magnitude (Fig.~\ref{fig:simu}, right). The discrepancy remains even when accounting for the uncertainty of sSFRs measured within a $<50$ kpc radius of the proto-BCGs in TNG300. This stark discrepancy poses a challenge to models of massive cluster galaxy formation in TNG300.
It is unclear why the TNG300 fails to reproduce the quiescence of massive proto-cluster galaxies. This could be caused by a combination of many effects.
At first, as suggested by the potential quenching mechanisms of the two quiescent galaxies, TNG300 might be   lacking  strong starburst and AGN feedback, and hence inefficient to quench star formations. This picture   agrees with a  recent study by \cite{Kimmig2023simu} for field galaxies.
Second, the SFHs of BCGs might   not be monotonically decreasing with cosmic time as the star formation of quiescent galaxies can be rejuvenated (e.g., \citealt{Remus2023simu}), and the quiescence on a timescale shorter than the time stamp spacing of TNG300 would be missed in the simulation.
Third, recent studies found that Illustris overpredicts the \cite{Madau2014a} SFR density by a factor of two at $z\sim3.5$ (e.g., \citealt{Shen2022TNG_SFRD}), which could partially overestimate the SFRs of BCGs.
Furthermore, the discrepancy could also be due to the different methods with which the SFR and $M_*$ are measured.
We note that the number density of quiescent cluster galaxies is a more straightforward quantity for comparison with the simulations; however, the sample size at $z>3$ is too small to give a good constraint on the number density. Fortunately, identifying a large sample of quiescent cluster galaxies at high redshift will come true soon with the Euclid telescope. Future work comparing a large sample with dedicated cluster simulations will be essential to solve the problem, for example Cluster-EAGLE \citep{Barnes2017ClusterEAGLE}, Magneticum \citep{Remus2023a_simu}, FLAMINGO \citep{Schaye2023FLAMINGO}, and TNG-Cluster \citep{Nelson2023TNG-Cluster}.

\section{Conclusions}

Using JWST and ancilary data in the EGS field, we discovered a large-scale structure, the Cosmic Vine, at $z=3.44$. Our conclusions are as follows:

1. Cosmic Vine is confirmed by 20 galaxies with spectroscopic redshifts at $3.43<z<3.45$. It hosts six galaxy overdensities of $\sim200$ candidate members in a vine-like structure extending over  $\sim4\times0.2$~pMpc$^2$.

2. The two most massive galaxies ($M_*\approx10^{10.9}~M_\odot$) in the  Cosmic Vine are found to be quiescent with bulge-dominated morphology. This unambiguously demonstrates that massive quiescent galaxies can form in growing large-scale structures at $z>3$, disfavoring the environmental quenching mechanisms that require a virialized cluster core.

3. We derived a halo mass of log$(M_{\rm halo}/M_\odot)=12.7$ for the primary overdensity peak in Cosmic Vine. Comparisons with simulations suggest that the Cosmic Vine would form a cluster with halo mass $M_{\rm halo}>10^{14}M_\odot$ at $z=0$, and the two massive galaxies are likely forming the BCGs.

4. In a comparison with the sSFR of proto-BCGs in the TNG300 simulation, we found that the observed sSFRs of massive quiescent galaxies in $z>3$ dense environments are significantly lower by one to two orders of magnitude. This stark discrepancy poses a potential challenge to the models of massive cluster galaxy formation.

A large sample of quiescent cluster galaxies at high redshift and comprehensive comparisons with dedicated cluster simulations will be essential to solving the discrepancy between observations and simulations, and will thereby shed  light on the detailed physics of cluster formation.

\begin{acknowledgements}
SJ acknowledges the financial support from the European Union's Horizon Europe research and innovation program under the Marie Sk\l{}odowska-Curie grant No. 101060888.
The Cosmic Dawn Center (DAWN) is funded by the Danish National Research Foundation under grant DNRF140.
GEM and SG acknowledge financial support from the Villum Young Investigator grant 37440 and 13160. APV and TRG acknowledges support from the Carlsberg Foundation (grant no CF20-0534). 
We acknowledge the CANDIDE cluster at the Institut d’Astrophysique de Paris, which was funded through grants from the PNCG, CNES, DIM-ACAV, and the Cosmic Dawn Center and maintained by Stephane Rouberol.

%XXX We also wish to acknowledge the following open source software packages used in the analysis: {Astropy \citep{astropy:2013, astropy:2018, astropy:2022}}, {Matplotlib \citep{Hunter:2007}}, {Numpy \citep{Harris2020_numpy}}, {Scipy \citep{2020SciPy-NMeth}}, TOPCAT \citep{Taylor2005Topcat}. 
\end{acknowledgements}

\bibliographystyle{aa}
\bibliography{biblio}

\begin{appendix}

\section{Supporting material}

\begin{figure}%
\centering
\includegraphics[width=0.47\textwidth]{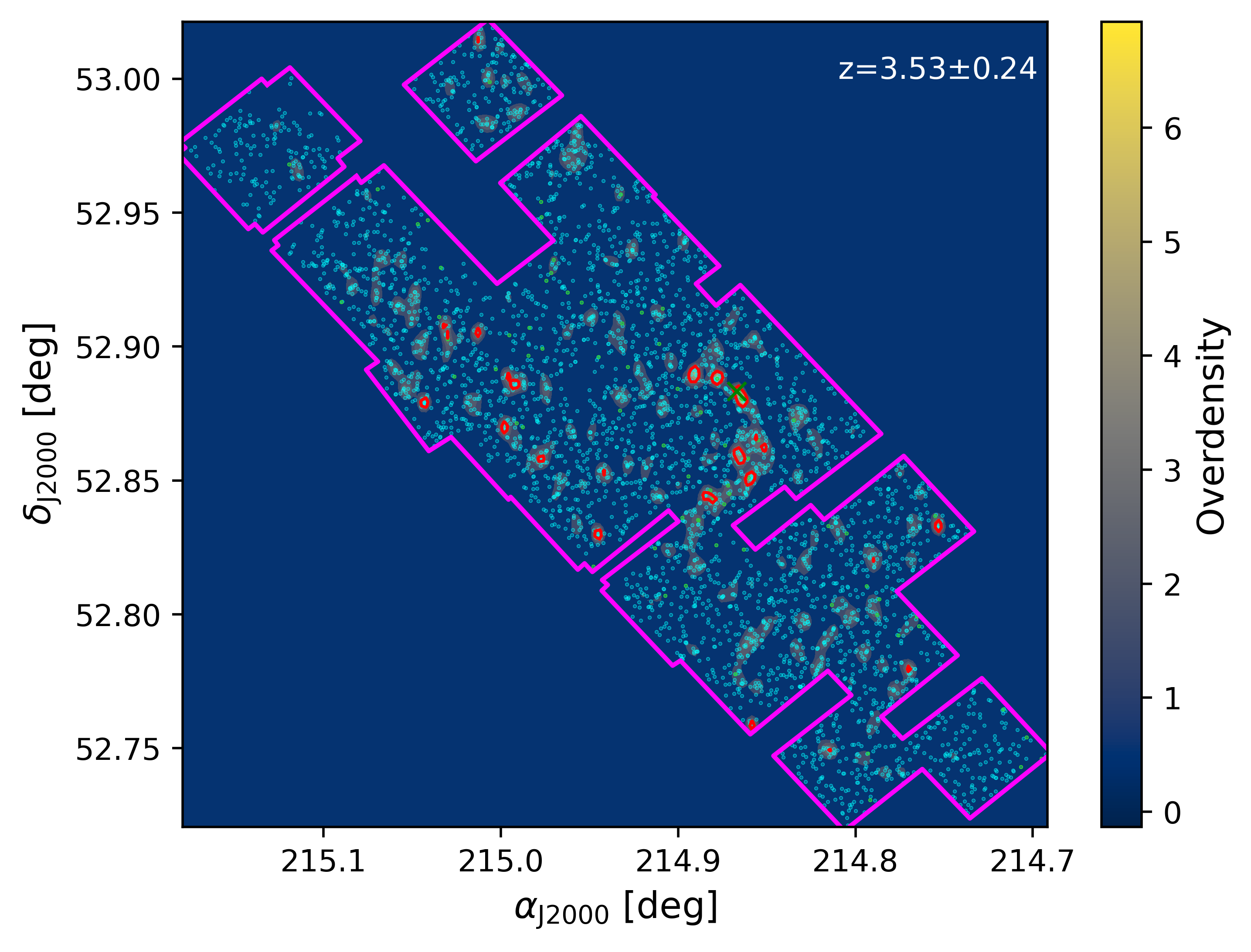}
\includegraphics[width=0.42\textwidth]{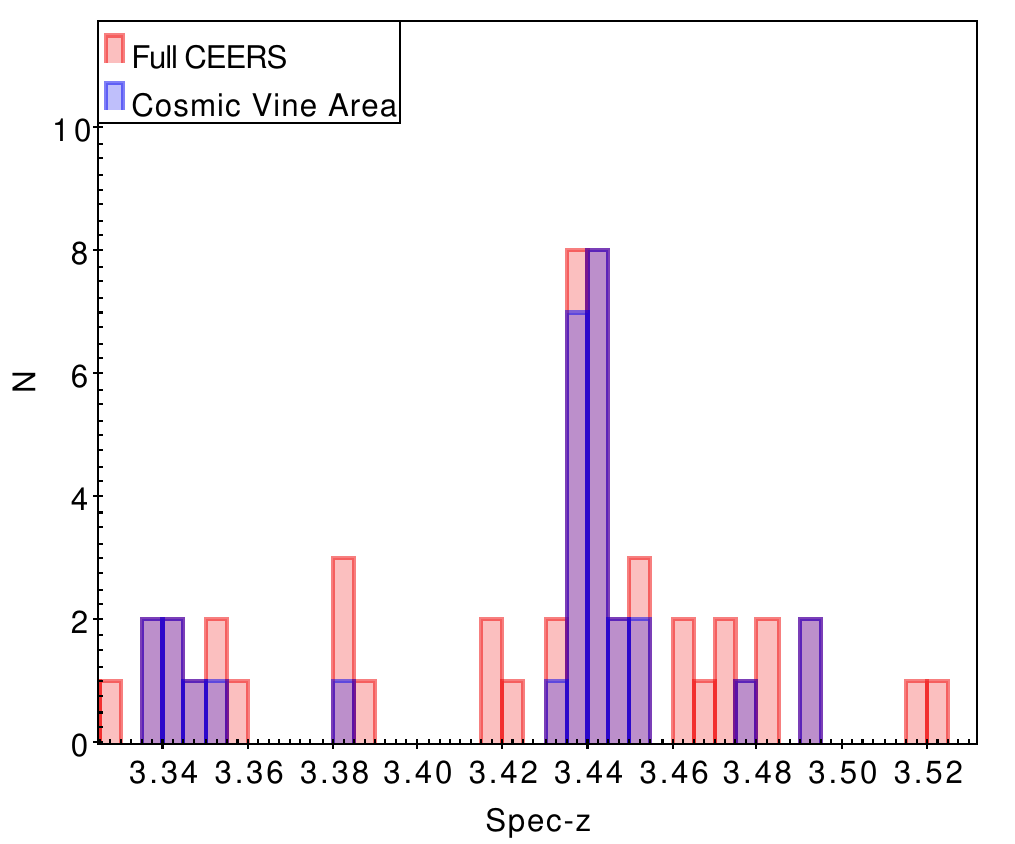}
\caption{Selection of the Cosmic Vine.
{\it Top:} Overdensity map in the CEERS field. The Cosmic Vine is highlighted with the 4$\sigma$ contour in red, and the green cross gives the position of Galaxy A.
{\it Bottom:} Spectral redshifts in the full CEERS field (pink) and the  Cosmic Vine area (purple).
}
\label{fig_selection}
\end{figure}

\begin{figure}%
\centering
\includegraphics[width=0.5\textwidth]{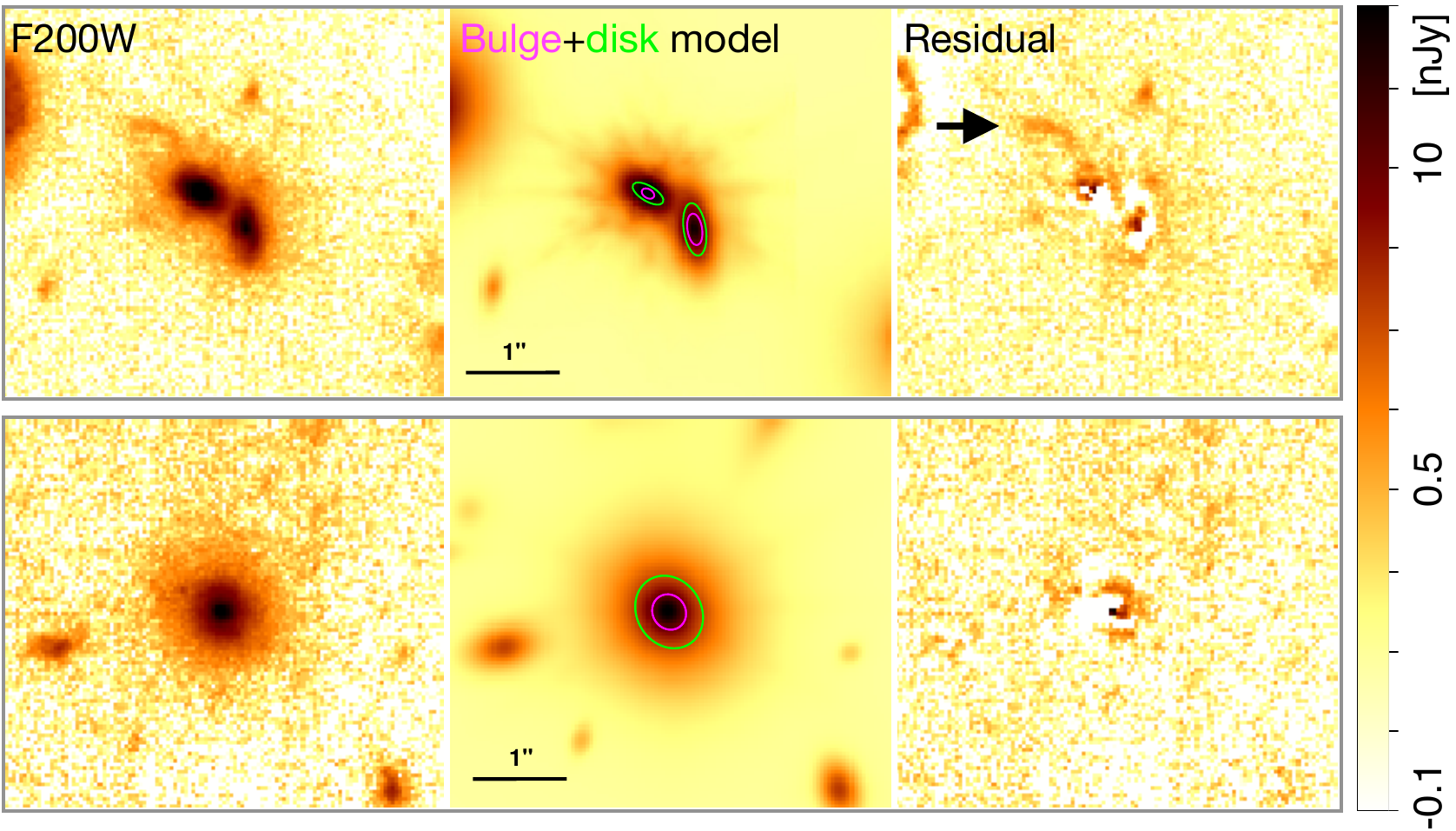}
\caption{Bulge+Disk decomposition in F200W image for Galaxy A+B (top) and E (bottom). We show the images, models, and residuals in log scale with identical limits. The magenta and green circles mark the effective radius of the bulge and disk models, respectively. The tidal tail is indicated by the  arrow.}
\label{fig_morph}
\end{figure}

\begin{figure}[ht]
\centering
\includegraphics[width=0.45\textwidth]{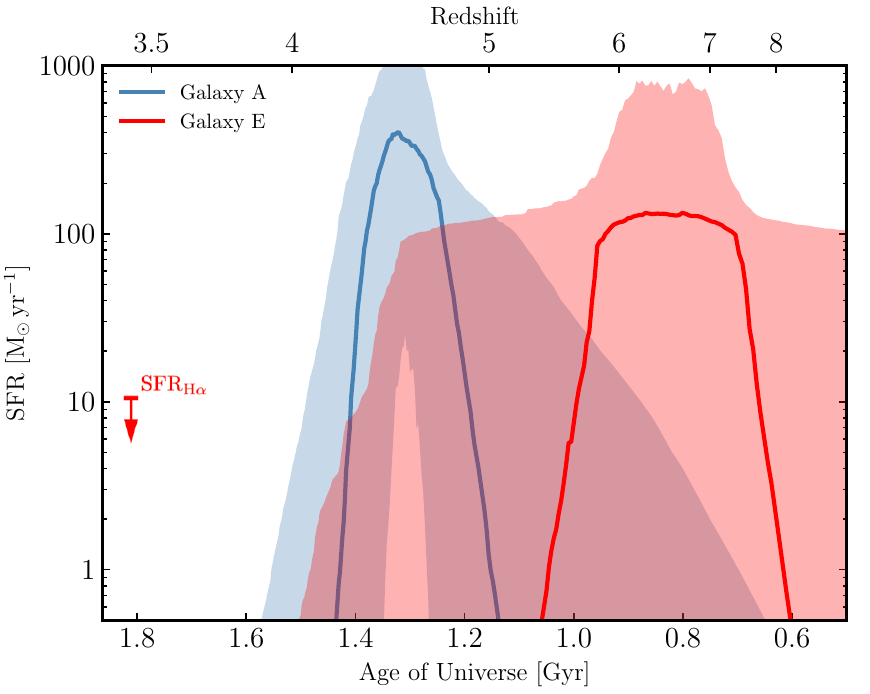}
\caption{SFHs of Galaxies A and E from Bagpipes fitting. The shaded area shows the 1$\sigma$ uncertainty. The red arrow gives the SFR upper limit derived from H$\alpha$ emission.
}
\label{fig_SFH}
\end{figure}

\begin{figure}[ht]
\centering
\includegraphics[width=0.45\textwidth]{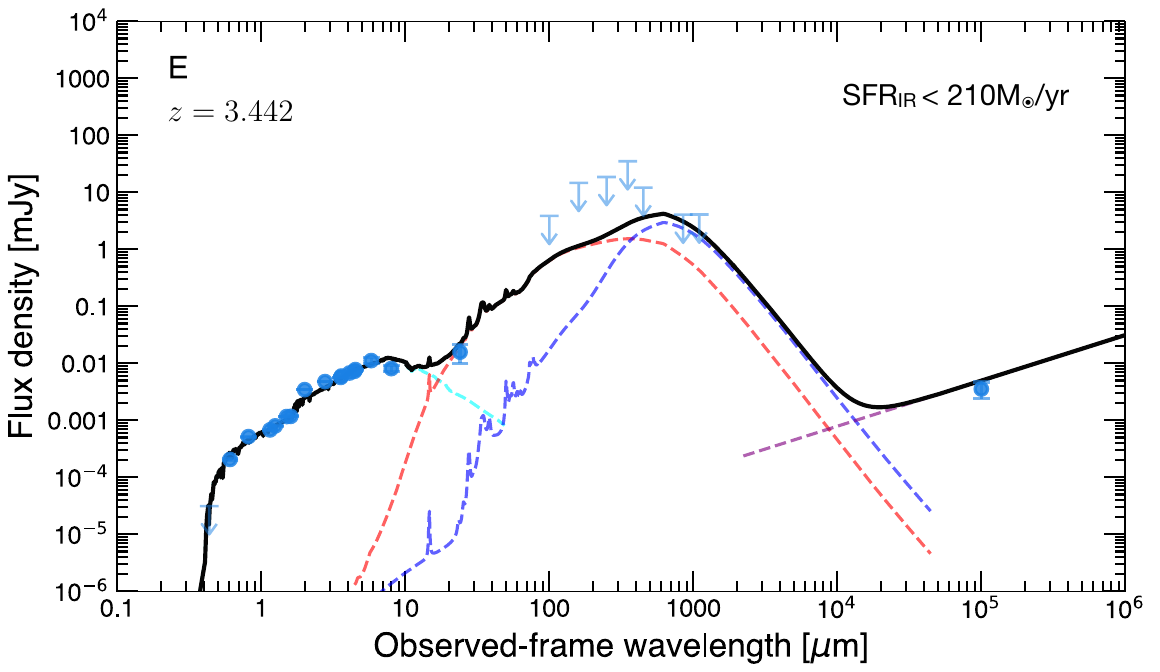}
\caption{NIR to radio SED of Galaxy E using MiChi2 \citep{Liu_DZ2021MiChi2}. From left to right, the dashed lines show the best-fit models of stellar (cyan, \citealt{BC03}), warm dust (red), cold dust (blue, \citealt{Magdis2012SED}), and an IR-correlated radio component (purple, \citealt{Magnelli2015}).
}
\label{fig_FIR}
\end{figure}

\end{appendix}

\end{document}